\documentclass[pre, twocolumn, superscriptaddress, floatfix, nofootinbib]{revtex4-1}  

\usepackage[paperwidth=210mm,paperheight=297mm,centering,hmargin=2cm,vmargin=2.5cm]{geometry}

\usepackage[pdftex]{graphicx}
\usepackage{amsmath,amsfonts,amssymb}
\usepackage{natbib}
\usepackage{MnSymbol}
\usepackage{csquotes}
\usepackage{lipsum}

\usepackage{hyperref}
\usepackage{xcolor} 
\usepackage{array}
\usepackage{pgfplots}
\usepackage{graphicx}
\usepackage{dsfont}
\usepackage{adjustbox}

\definecolor{bluemoi}{rgb}{0.25,0.50 ,0.75} 

\hypersetup{
	bookmarksopen=true,
	pdftitle=" ",
	pdfauthor=" ", 
	pdfsubject=" ", 
	pdftoolbar=true, 
	pdfmenubar=true, 
	pdfhighlight=/O, 
	colorlinks=true, 
	pdfpagemode=UseNone, 
	pdfpagelayout=SinglePage, 
	pdffitwindow=true, 
	linkcolor=bluemoi, 
	citecolor=bluemoi, 
	urlcolor=bluemoi 
}

\makeatletter
\renewcommand{\figurename}{\sf \textbf{Figure}}
\renewcommand{\thefigure}{\arabic{figure}}
\renewcommand{\fnum@figure}{\sf\textbf{\figurename}~\textbf{\thefigure}}
\renewcommand{\tablename}{\sf\textbf{Table}}
\renewcommand{\thetable}{\arabic{table}}
\renewcommand{\fnum@table}{\sf\textbf{\tablename}~\textbf{\thetable}}
\makeatother

\makeatletter
\renewcommand\@biblabel[1]{#1} 

\renewcommand\newblock{\hskip .11em\@plus.33em\@minus.07em}
\makeatother

\setcitestyle{authoryear,round}
\bibpunct{(}{)}{;}{a}{,}{}

\makeatletter
\newcommand{\removeperiod}{\@ifnextchar.{\@gobble}\relax}
\makeatother

\begin{document}
	
\title{Intersectional approach of everyday geography} 

\author{Julie Vall\'ee}
\thanks{Julie Vallée (julie.vallee@parisgeo.cnrs.fr) and Maxime Lenormand (maxime.lenormand@inrae.fr) contributed equally to this work.}
\affiliation{G\'eographie-cit\'es (CNRS, Univ Paris 1 Panthéon-Sorbonne, Univ Paris Cité, EHESS), Paris-Aubervilliers, France}
	
\author{Maxime Lenormand}
\thanks{Julie Vallée (julie.vallee@parisgeo.cnrs.fr) and Maxime Lenormand (maxime.lenormand@inrae.fr) contributed equally to this work.}
\affiliation{TETIS (Univ Montpellier, AgroParisTech, Cirad, CNRS, INRAE), Montpellier, France}

\begin{abstract} 
Hour-by-hour variations in spatial distribution of gender, age and social class within cities remain poorly explored and combined in the segregation literature mainly centered on home places from a single social dimension. Taking advantage of 49 mobility surveys compiled together (385,000 respondents and 1,711,000 trips) and covering 60\% of France's population, we consider variations in hourly populations of 2,572 districts after disaggregating population across gender, age and education level. We first isolate five district hourly profiles (two ’daytime attractive’, two ’nighttime attractive’ and one more ’stable’) with very unequal distributions according to urban gradient but also to social groups. We then explore the intersectional forms of these everyday geographies. Taking as reference the dominant groups (men, middle-age and high educated people) known as concentrating hegemonic power and capital, we analyze specifically whether district hourly profiles of dominant groups diverge from those of the others groups. It is especially in the areas exhibiting strong increase or strong decrease of ambient population during the day that district hourly profiles not only combine the largest dissimilarities all together across gender, age and education level but are also widely more synchronous between dominant groups than between non-dominant groups (women, elderly and low educated people). These intersectional patterns shed new light on areas where peers are synchronously located over the 24-hour period and thus potentially in better position to interact and to defend their common interests.

\end{abstract}

\maketitle
\section*{Introduction}
	
People's daily mobility in cities has been largely explored with particular attention to differences in trips number, distances traveled and transportation modes according to gender, age and social class. However it remains much less
common to examine how people's daily mobility leads to hourly variations in spatial concentration of gender, age and social subgroups within cities. Such a blind spot is unfortunate because spatial and temporal constrains that demographic and social subgroups unequally encounter in their daily life do not only translate into differences in the ways they move around but generate also divergent spatial patterns in their co-locations and in segregation of cities around the clock.

Social and political determination of domestic, work, and leisure places and times shape hourly rhythms both of social groups and of places. While the inequitable positioning of women in society structures gendered utilisation of urban space \citep{Cresswell_2008}, co-locations of women and men in cities throughout the day remain understudied in segregation literature probably because the traditional model of family and heterosexual couple induces gender parity in housing, at least at night, and by extension in residential areas. One can yet think that the gender mix observed in residential areas decreases during the day depending on the type and location of daily activities (professional but not only) with some neighborhoods becoming overwhelmingly female or male \citep{Vallee_2020} in link with the differential presence and involvement of men and women in public and private spheres \citep{McDowell_1999}. This is also true for the co-locations of different age groups, whose residential patterns \citep{Cowgill_1978, LaGory_1981, Winkler_2012} can greatly change throughout the day \citep{Abbasi_2021}. As developed in Lefebvre's Rhythmanalysis \citep{Lefebvre_1992}, powerful groups imprint rhythms upon places. By scheduling public transportation or events in public spaces, by planning land use such as workplaces, shopping and recreational facilities, public and private actors may assemble, frame and co-ordinate population daily locations. Such strategies of synchronization play an important part in the production and the control of population in places. Some inclusive night-time areas can then become exclusive for some subgroups of population during the day, and inversely \citep{Karrholm_2009, Vallee_2017}. Focusing on socioeconomic position or ethnicity, some studies have explored how social segregation vary around the clock in cities such as Miami \citep{Wong_2011}, Tallin \citep{Silm_2014}, Detroit \citep{Farber_2015},  Paris \citep{LeRoux_2017} or Atlanta \citep{Park_2018}. However, these rare studies of everyday segregation focus solely on a single axe of difference (e.g. income) and do not consider everyday exclusion process impacting all together gender, age and socioeconomic position from an intersectional point of view.

Following the mobility turn, or new mobilities paradigm \citep{Sheller_2006}, there is a need to bear on connections between daily (im)mobility and the intersectional forms of space-time experiences. We aim here to provide empirical keys on the way subordinates deal with ‘multi-layered and routinized forms of domination’ \citep{Crenshaw_1991} in their space-time experiences and to fill a gap in segregation literature where reference to intersectionality remains curiously absent and, at best, mentioned only in passing, as recently underlined by Hopkins \citep{Hopkins_2019}. Such intersectional approach of everyday geography would be useful to emphasize spatial injustice in daily access to the power structure. Finally, unlike the rare studies about everyday segregation which are mainly limited in geographical scope \citep{Muurisepp_2022}, the present paper concerns a large sample of districts and city regions allowing comparisons across city regions or across similar urban rings from different city regions.

To date, studies exploring daily people's locations, ambient populations and everyday segregation have relied on different types \citep{Whipp_2021, Muurisepp_2022}. The more recent one refer to the digital traces that people leave with their mobile phones \citep{Lenormand_2015, OlteanuRaimond_2012, Jiang_2016, Song2010, Hanaoka_2018}, their transportation cards \citep{Zhong_2016}, their credit cards \citep{Montjoye_2015, Louail_2017} or their participation in social media - such as Twitter \citep{Wang_2018,Heine_2021}. Digital traces depend on the operator market share and may suffer from representative issues in users' demographics. With phone traces, scholars are also inclined to restrict analysis to inner cities where the grid of mobile phone towers is sufficiently small to enable accurate analyses \citep{Sakarovitch2019}. Moreover, in digital data person-based information such as ethnicity or social class (and also sex and age) are very often lacking for confidential/privacy reasons, even if there are some exceptions \citep{Lenormand_2015_demo, Gauvin_2020}. To overcome the lack of individual social information in digital data, some recent studies use the aggregated social profile of residential areas from census or fiscal databases to infer users social class or race and enrich the data. This residential and ecological inference is yet ineffective to extract some individual information (i) about social groups that are heterogeneous located in residential areas (e.g. middle classes, gender or age) or (ii) in areas where density of mobile phone towers is low (e.g. rural areas) and where larger census tracts often gather more heterogeneous population. Besides digital traces, a more traditional literature use census data to examine geographies of segregation at home and at work \citep{ellis_2004, Hellerstein_2008}. However, these home-work data exclude the large part of the population who is non-active (unemployed, retired, at home, etc.) and do not consider locations experienced through leisure, errand activities, or visits to friends and family. Moreover, census data give no information about the temporal agenda inhibiting space-time analysis of everyday geography. A third data source is also available in some specific countries and cities: the origin-destination surveys (often based on travel diaries). They provides some precise space-time data including a large range of person-based data (e.g. sex, age, education level, occupational status) from a representative population. Often seen as less fashionable that digital traces, with smaller sample of respondents, over a more limited time span, or with errors due to self-reporting,  these traditional data sources are yet valuable especially when they concern a large number of cities and ensure that observations made in one city remain valid for other cities \citep{Cottineau_2019}. In France, where origin-destination surveys are carried out in standardized way in a vast range of city regions, microdata about daily trips can be compiled to concern 385,000 respondents across 2,572 districts located in 49 city regions \citep{mobiliscope_datav4-1, Cerema}. This dataset provide a large sample bearing comparison with studies using mobile-phone \citep{Song2010, Phithakkitnukoon_2012} or socio-media data \citep{Wang_2018} for which initial and massive samples get often drastically reduced after data cleaning.

Instead of using these French microdata to analyze daily \textit{trips}, we focus here on people's hourly \textit{locations} to explore in everyone of the 2,572 districts the number of present population of each social group (defined from their gender, age or education level) across the 24-hour period. Actually, these district hourly profiles are at the heart of our empirical analysis: we use them to compare the everyday geography of each social group.

By exploring differences in district hourly profiles (called mismatch in the present paper), we aim to reveal the intersectional forms of everyday exclusion resulting from the spatial and temporal constraints that social groups unequally encounter in the course of their daily life combined with their residential segregation. Our argument is as follows: if hourly profiles of different social groups in a same district are found to perfectly overlap, it means that the district has similar hourly attractiveness whatever the social group under consideration and that every social group equally access to it over the 24-hour period. Conversely, if district hourly profiles of different social groups do not overlap (i.e. exhibit a mismatch), it means that the district under consideration has uneven hourly attractiveness and that its everyday use is different from one social group to another. Taking into consideration social subgroups known as concentrating hegemonic power and capital \citep{Bessiere_Gollac_2020, Atkinson_1971, Garbinti_2018}, we define men, middle-age and high educated people as belonging to dominant groups. By contrast, we defined women, elderly people and low-educated people as belonging to non-dominant groups. Even if age is more rarely considered in intersectional studies, we define elderly people as belonging to non-dominant groups according to the stigma and exclusion attendant to old age in the contemporary societies of the global North and to the ways this stigma vary with gender and social class \citep{Calasanti_2015}. Here, we explore whether (and where) district hourly profiles of dominant groups diverge from those of the others groups. More specifically, the present paper aims to meet the following 4 objectives: (1) describe district hourly profiles and differences in their distribution according to social groups but also according to district location; (2) measure to what extent a same district exhibits mismatch in its hourly profiles according to gender, age or educational groups under consideration; (3) highlight districts combining all together large mismatch in their hourly profiles across gender, age and education; (4) explore mismatch in district hourly profiles between dominant groups and explore whether (and where) they differ from mismatch in district hourly profiles between non-dominant groups.
	
\section*{Data}

\subsection*{Initial trip dataset} 

We used microdata coming from large origin-destination surveys conducted from 2009 to 2019 (during the autumn and winter seasons) in 49 French city regions covering 39.5 millions of inhabitants (i.e. 60\% of France's population). The sample contains 385,000 respondents aged 16 years and more and questioned about all their trips (place and time of departure/arrival for every activity) the day before of the survey (from Monday to Friday). Additional person-based information regarding gender, age and educational level of the respondents are also provided in the surveys (but no ethnicity or race information is collected in accordance with French state’s policy rejecting any references to racial or ethnic minorities). We considered four age groups (16-24 years, 25-34 years, 35-64 years and 65 and more) and merged respondents’ achieved level of education in four groups of individual educational status: low (middle school or less), middle-low (high school without Baccalaur{\'e}at\footnote{French high school diploma}), middle-high (up to two years after Baccalaur{\'e}at) and high (three years or more after Baccalaur{\'e}at). Sample contains 385,000 respondents and 1,711,000 trips across 49 city regions (see Table S1 in Supplementary Material for summary statistics).

\subsection*{Hourly location dataset} 

To understand how the hourly distribution of individuals changes over spatial and social dimensions, trip dataset was transformed into hourly location dataset in which every location was defined at district level. Districts are the smallest units in which it is relevant to aggregate data when it comes to not only ensuring sufficient sample size for statistical analysis but also protecting confidentiality of personal data for the provision of open data (see below). Districts correspond to a neighborhood in core urban areas and to a municipality or a group of municipalities in peripheral areas. 24 hourly time steps were defined to get 24 cross-sectional pictures of respondents' locations at exact hours (04.00, 05.00 etc.). Short locations in the interval between two exact hours have then not been registered in district hourly dataset. Transportation periods were also not considered except if respondents reported to use an 'adherent' mode of transportation (i.e. walking or cycling). In this case, half of the trip was considered as located in the district of origin and the other half as located in the district of destination \citep{LeRoux_2017}. In the (rares) cases where 'adherent' trip  symmetrically straddling an exact hour, location at this very hour was chosen to be in the district where respondent stayed the shortest time (because the longest duration taking place in the other district has a high probability to be registered at another hour).

\begin{figure}[!h]
	\centering 
	\includegraphics[width=8cm]{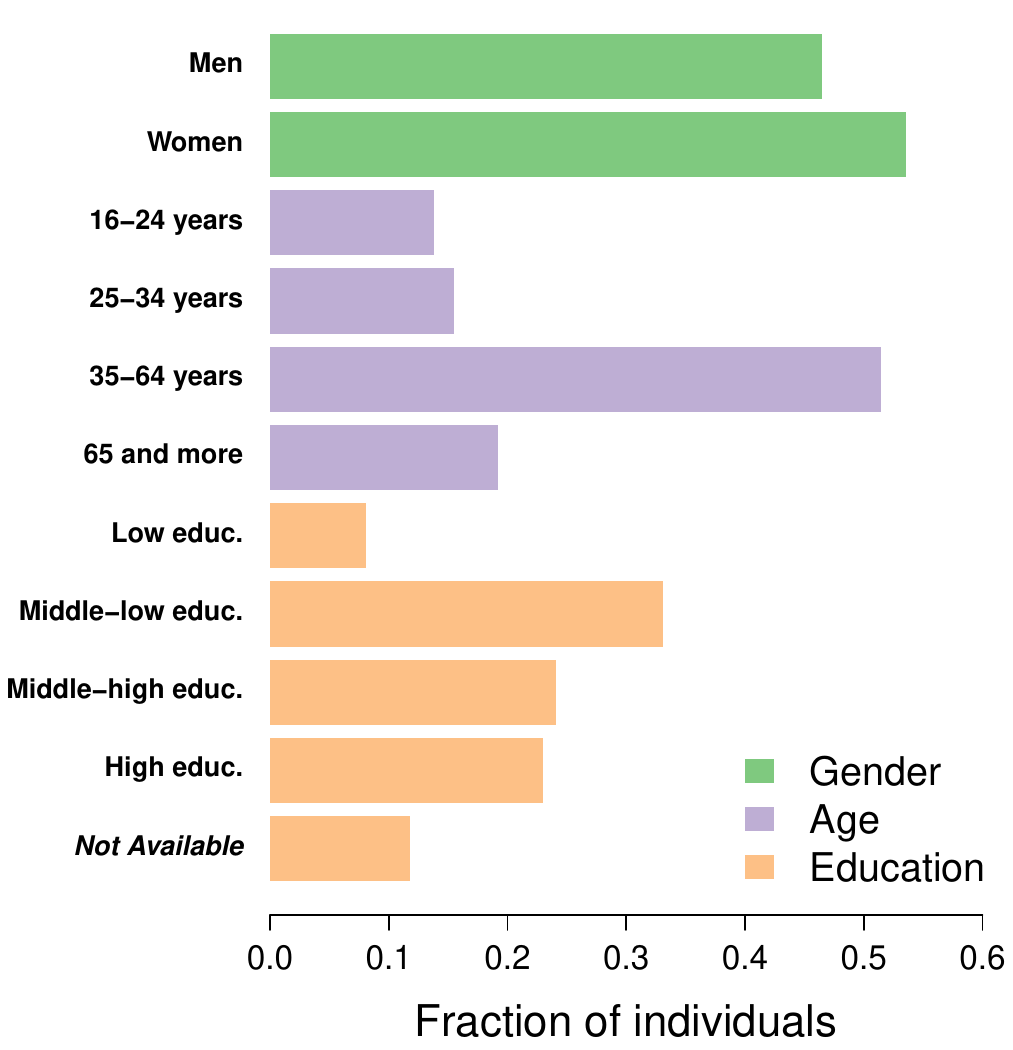}
	\caption{\textbf{Fraction of individuals (after the weighting process) according to gender, age and educational groups.}
		\label{Fig1}}
\end{figure}

In everyone of the 2,572 districts the number of present population over the 24-hour period was estimated for each of the 10 population groups under consideration: men or women; 16-24 years, 25-34 years, 35-64 years or 65 years and more; low, middle-low, middle-high or high education. These estimations took into account the respondent's weighting coefficients provided in the origin-destination surveys \citep{Cerema}. Figure \ref{Fig1} displays the fraction of individuals under consideration (after the weighting process) according to gender, age and educational groups. We can note that 12\% of people have no available information about their achieved level of education: it concerns mainly (69\%) the youngest people (16-24 yrs.) because many of them are still in school. 

\begin{figure}[!h]
	\centering 
	\includegraphics[width=\linewidth]{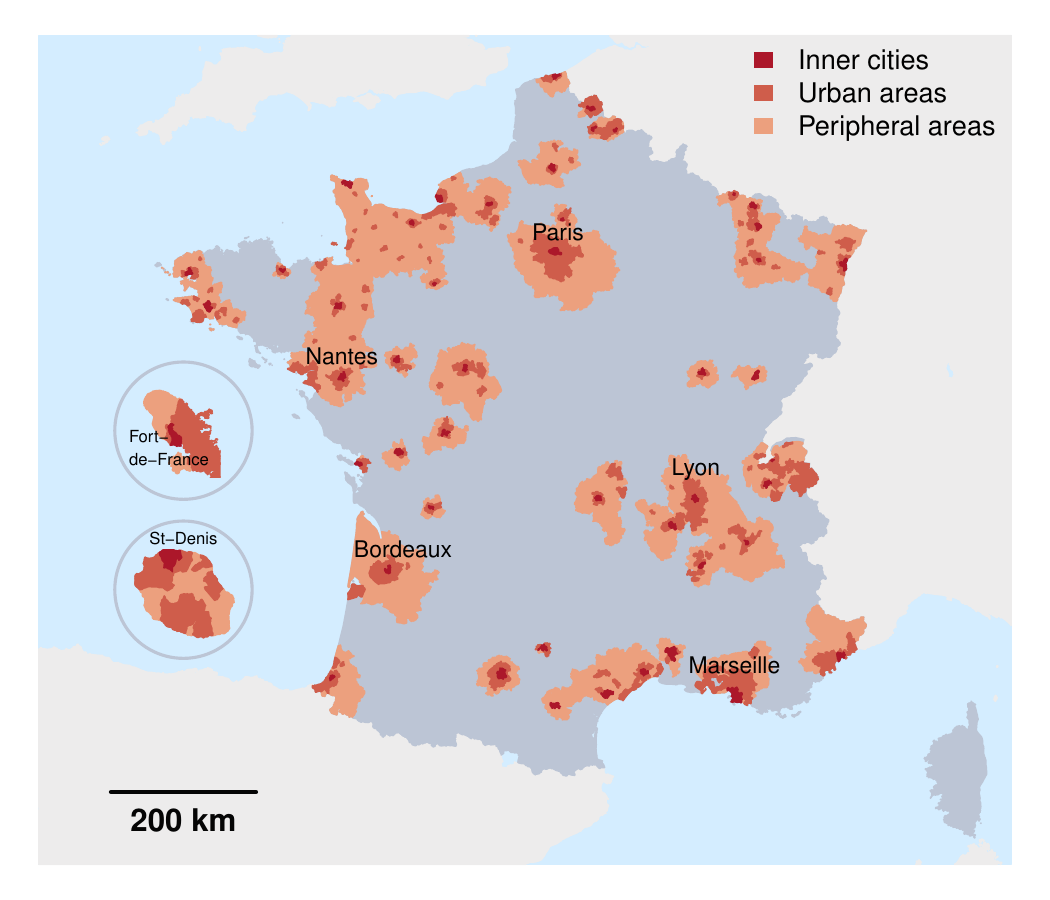}
	\caption{\textbf{Map of the studied areas.} The dataset is composed of 2,572 districts gathered in 49 French city regions. The two insets concern Martinique (top) and La Réunion (bottom). The districts are divided in three categories: inner cities, urban areas and peripheral areas.
		\label{Fig2}}
\end{figure}

Located in 49 French city regions, these 2,572 districts can be defined according to urban gradient (Figure \ref{Fig2}): 653 districts (25.5\%) belong to the inner areas of the 49 city regions, 1,074 districts (41.9\%) are located outside inner cities but in urban areas (major, medium or small urban poles), and 834 districts (32.6\%) are part of peripheral areas (corresponding to surroundings of urban poles, multi-polarized or isolated areas) as defined in the French functional urban zoning ('Zonage en Aires urbaines') proposed by the French National Statistical Institute (Insee).

\subsection*{Temporal signal normalization} 

For each city region $c$, the distribution of the different social groups in space and time was quantified with a three-dimensional matrix $N^c=(N^c_{i,j,h})$ representing the number of individuals belonging to the social group $i \in |[1,m]|$ present in district $j \in |[1,n]|$ at hour $h\in |[1,24]|$ during a typical weekday. We considered here $m=11$ overlapping social groups: the total population (hereafter called 'All'), the gender (men or women), the four age groups and the four educational groups. 

We first normalized the three-dimensional matrix by the total hourly activity of the social group $i$ in each city region $c$ (Equation \ref{Normalization1}). 
\begin{equation}
\hat{N^c}_{i,j,h} = \frac{\displaystyle N^c_{i,j,h}}{\sum_{k=1}^{n} \displaystyle N^c_{i,k,h}}
\label{Normalization1}
\end{equation}
This normalization was made to ensure that the total number of individuals belonging to a given social group present in a city region was constant along the day. We further normalized this quantity to obtain a temporal signal summing to 1 for a given social group and a given district (Equation \ref{Normalization2}). %
\begin{equation}
T^c_{i,j,h} = \frac{\displaystyle \hat{N^c}_{i,j,h}}{\sum_{k=1}^{24} \displaystyle \hat{N^c}_{i,j,k}}
\label{Normalization2}
\end{equation}
At the end of the process, we obtained a final sample of 28,281 temporal signals reflecting the temporal distribution of $11$ social groups in the 2,572 districts (hereafter called sociodistricts). Eleven sociodistricts without activity (2 for low educated group and 9 for high educated group) were removed.

\section*{Methods}

\subsection*{District hourly profiles of social groups}

To group together sociodistricts exhibiting similar temporal signals, we performed an ascending hierarchical clustering using Ward's metric and Euclidean distances as agglomeration method and dissimilarity metric, respectively \citep{Hastie_2009}. From this clustering we isolated different district hourly profiles and explored whether they were unequally distributed according to social groups, urban gradient and city regions (\textit{objective 1}). Chi-square tests were used to determine whether there was a statistically significant difference between distributions (with significance level of 0.01).

\subsection* {Mismatch in district hourly profiles}

To highlight to what extent a district gather different hourly profiles according to social groups, we computed - for every pair of social groups - a mismatch index defined as the Euclidean distance between their two hourly profiles in every district under consideration. This Euclidean distance was normalized by the maximum reachable distance between the two extreme hourly profiles. We privileged this indicator because of its direct interpretation. Indeed, for a given district, the mismatch between two different social groups varies from 0, when the two social groups have exactly the same profile, to 1, when the two social groups have opposite profiles with the maximum reachable Euclidean between profiles.

Although this mismatch index could be computed between any pairs of social group for a given district, we first computed mismatch in hourly profiles across each of the three sociodemographic variables (gender, age or education), i.e. 13 mismatch values per district (one for the gender, six for the age and six for the education). We then explored how mismatch indices differed for each of the three sociodemographic variables not only among the whole district sample (i.e. the 2,561 districts with full information), but also among subsamples of districts according to their location within city regions (i.e. urban gradient) and the city region they belong to (\textit{objective 2}). Notched boxplots were used to establish visually the significance of differences between mismatch values. The width of a notch displays the 95\% confidence interval information for the median: when the notches do not overlap, it provides evidence of a statistically significant difference between medians values.

\subsection* {Intersectional mismatch in district hourly profiles}

To extend analysis of mismatch in district hourly profiles with an intersectional lens, we explored more specifically whether district hourly profiles of dominant groups diverge from those of the others groups. Taking into consideration groups known as concentrating hegemonically power and capital \citep{Bessiere_Gollac_2020, Atkinson_1971, Garbinti_2018}, we defined \emph{a priori} a 'dominant' group for each of the three sociodemographic variables: men for the gender, the 35-64 years old group for the age and the high educational level for the education.

In a first step, we focused on the seven mismatch indices measuring differences in district hourly profiles across gender, age and education using dominant groups as a reference (one mismatch value for gender, three for age and three for education). After having described one by one the gender-based mismatch, the age-based mismatch and the education-based mismatch, we aimed at highlighting districts combining large mismatches all together across gender, age and education (\textit{objective 3}). To provide some answers to this question, we relied here again on the Ward's metric and Euclidean distances to cluster together districts exhibiting a similar set of the seven mismatch values under consideration. In order to emphasize differences between the clusters, we based our cluster analysis on the concept of \textit{V-test} measuring the gap between the average mismatch within a cluster of districts and the average mismatch in the whole district sample by taking into account the mismatch empirical variance. For a given mismatch $X$ (mismatch between women and men for example) the V-Test defined in Equation \ref{vtest} was used to compare the average mismatch $\bar{X}_g$ measured in a cluster $g$ (with $n_g$ districts) with the average mismatch $\bar{X}$ in the whole sample ($n$ districts). We note that the denominator is the standard error of the mean in the case of a sampling without replacement of $n_g$ elements among $n$.
\begin{equation}
VT = \frac{\displaystyle \bar{X}_g - \bar{X}}{\displaystyle \sqrt{\frac{n-n_g}{n-1}\frac{\sigma^2}{n_g}}}
\label{vtest}
\end{equation}
where $\sigma^2$ stands for the empirical variance of the mismatch. Chi-square tests (with significance level of 0.01) were once again used to determine whether there was statistically significant differences in cluster distributions according to the districts' geographical location.

In a second step, we focused on differences in district hourly profiles \textit{between} dominant groups (men, middle-age and high educated people). We computed for every district the average of the three associated mismatch values (men vs. middle-age people; men vs. high educated people; middle-age vs. high educated people). This average mismatch value is called dominants' mismatch. For comparison purpose, we also considered the three mismatch values measuring differences in district hourly profiles \textit{between} non-dominant groups (women, elderly and low educated people) and computed similarly for every district the average mismatch in their hourly profiles (called non-dominants' mismatch) from the three associated mismatch values (women vs. elderly people; women vs. low educated people; elderly vs. low educated people). By comparing dominants' mismatch and non-dominants' mismatch values, we aimed at exploring whether synchronization in district hourly profiles \emph{between} dominant groups differed from synchronization in district hourly profiles \emph{between} non-dominant groups, and whether synchronization patterns were similar whatever the districts (\textit{objective 4}).

\subsection*{Data and materials availability} 

All data, code, and materials used in the present paper are available in open access. Initial datasets with hourly populations estimations of the 2,572 French districts come from \textit{Mobiliscope} (\url{www.mobiliscope.cnrs.fr}), an open interactive geovizualisation platform to explore cities around the clock \citep{mobiliscope_software}. Derived from version v4.1 of Mobiliscope, initial datasets are also stored under ODbL license in an open-access repository (\url{www.doi.org/10.5281/zenodo.7738571}). Procedures are made available in a public repository under License GPLv3 (\url{https://gitlab.huma-num.fr/mobiliscope/intersectionality}) and findings can be fully explored in a open dedicated cartographic platform (\url{https://shiny.umr-tetis.fr/Intersectionality}).

\section*{Results}

\subsection*{District hourly profiles of social groups}

\begin{figure*}[!ht]
	\centering 
	\includegraphics[width=\linewidth]{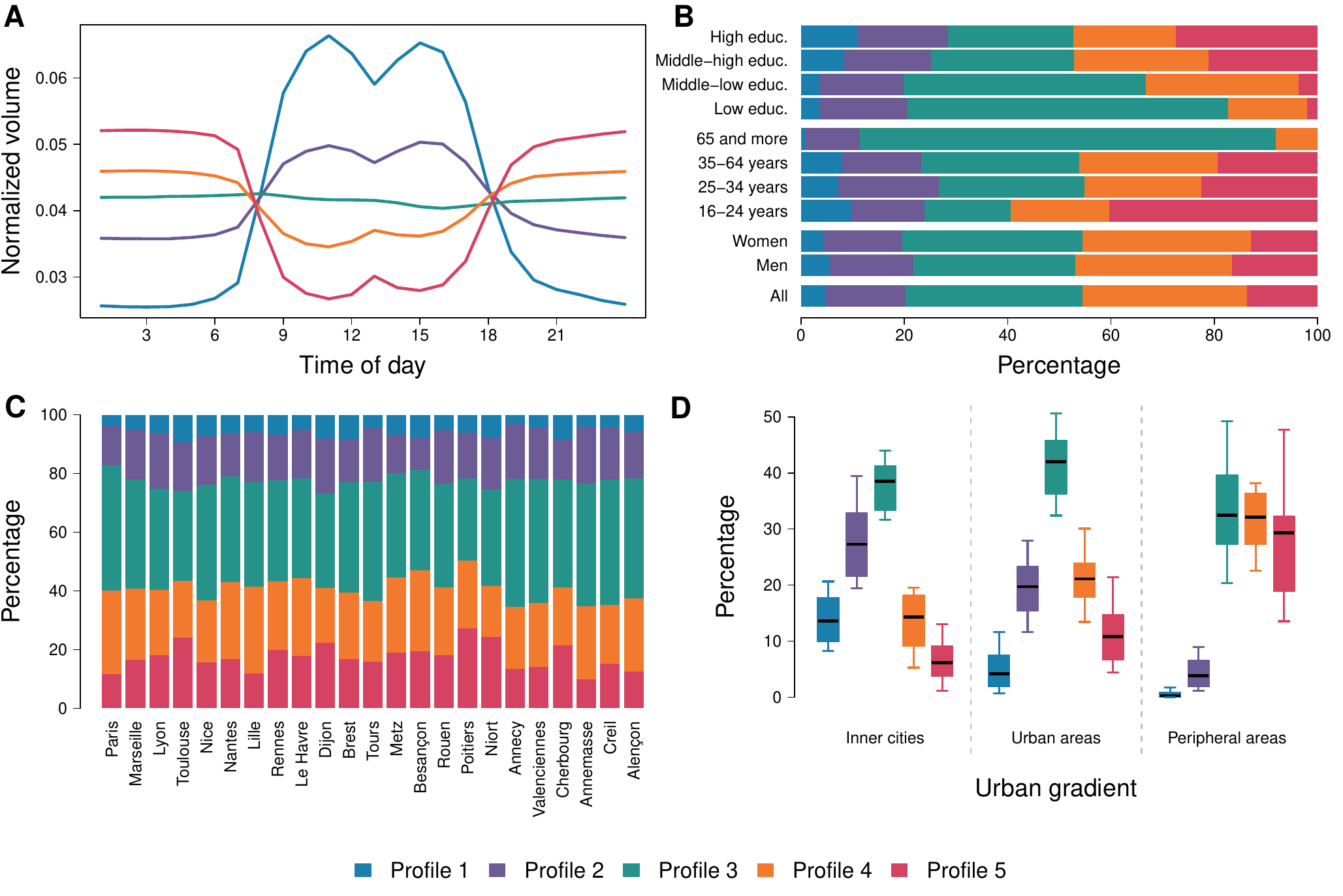}
	\caption{\textbf{District hourly profiles.} (A) Average temporal signals representing the five profiles. (B) Percentage of sociodistricts belonging to each profile according to the social groups and the total population (last line 'All'). (C) Percentage of sociodistricts belonging to each profile according to the city region (only 23 city regions in mainland France are displayed here: the six more populated and also those with the more recent data). (D) Boxplots of the percentage of sociodistricts per city region according to the profile and the urban gradient. Each boxplot is composed of the first decile, the first quartile, the median, the third quartile and the ninth decile.
		\label{Fig3}}
\end{figure*}

From ascending hierarchical clustering, we obtained five main profiles of sociodistricts exhibiting similar temporal signals (see Figures S1 and S2 in Supplementary Material for more information about clustering outputs). Figure \ref{Fig3}A shows how people concentration looks like for each of the five hourly profiles and Figure \ref{Fig3}B displays the distribution of the five hourly profiles. 

Two 'daytime attractive' profiles gather $22\%$ of sociodistricts ($6\%$ for the profile 1 and $16\%$ for the profile 2). They both show an increase of people concentration between 6am and 6pm with a small drop at lunchtime. The increase during the day is much greater for profile 1 than for profile 2. Two 'nighttime attractive' profiles gather $40\%$ of sociodistricts ($24\%$ for profile 4 and $16\%$ for profile 5) and exhibit conversely a decrease of people concentration from 6 am, a small increase of activity around 12pm and a decrease from 3pm. The decrease during the day is much greater for profile 5 than for profile 4. The temporal concentration patterns associated with profiles 1 and 2 appear to be the opposite of the ones associated with profiles 5 and 4, respectively. A last 'stable' behaviour (profile 3) gathering $38\%$ of sociodistricts exhibit similar people concentration over the 24 hour period, even if there is a very small decrease from 7am to 6pm. 

Distribution of district hourly profiles according to the social groups (Figure \ref{Fig3}B) highlights large and significant differences across gender, age and educational groups. The two attractive daytime profiles (profiles 1 and 2) are more common when educational level increases: for the high educational group there are about 11\% of their district hourly signals belonging to profile 1 compared to 4\% for the low educational group. Attractive daytime profiles are also less common when age decreases: for the young adults (16-24 years) about 10\% of their district hourly signals belong to profile 1 compared to 1\% for people aged 65 years. For gender, attractive daytime profiles are for example slightly more common for men than for women (5.4\% vs. 4.4\%, respectively for profile 1). Other patterns can be highlighted: districts exhibiting stable profile during the day (profile 3) are more numerous as age increases, as education level decreases and for women than for men.

\begin{figure*}[!ht]
	\centering 
	\includegraphics[width=\linewidth]{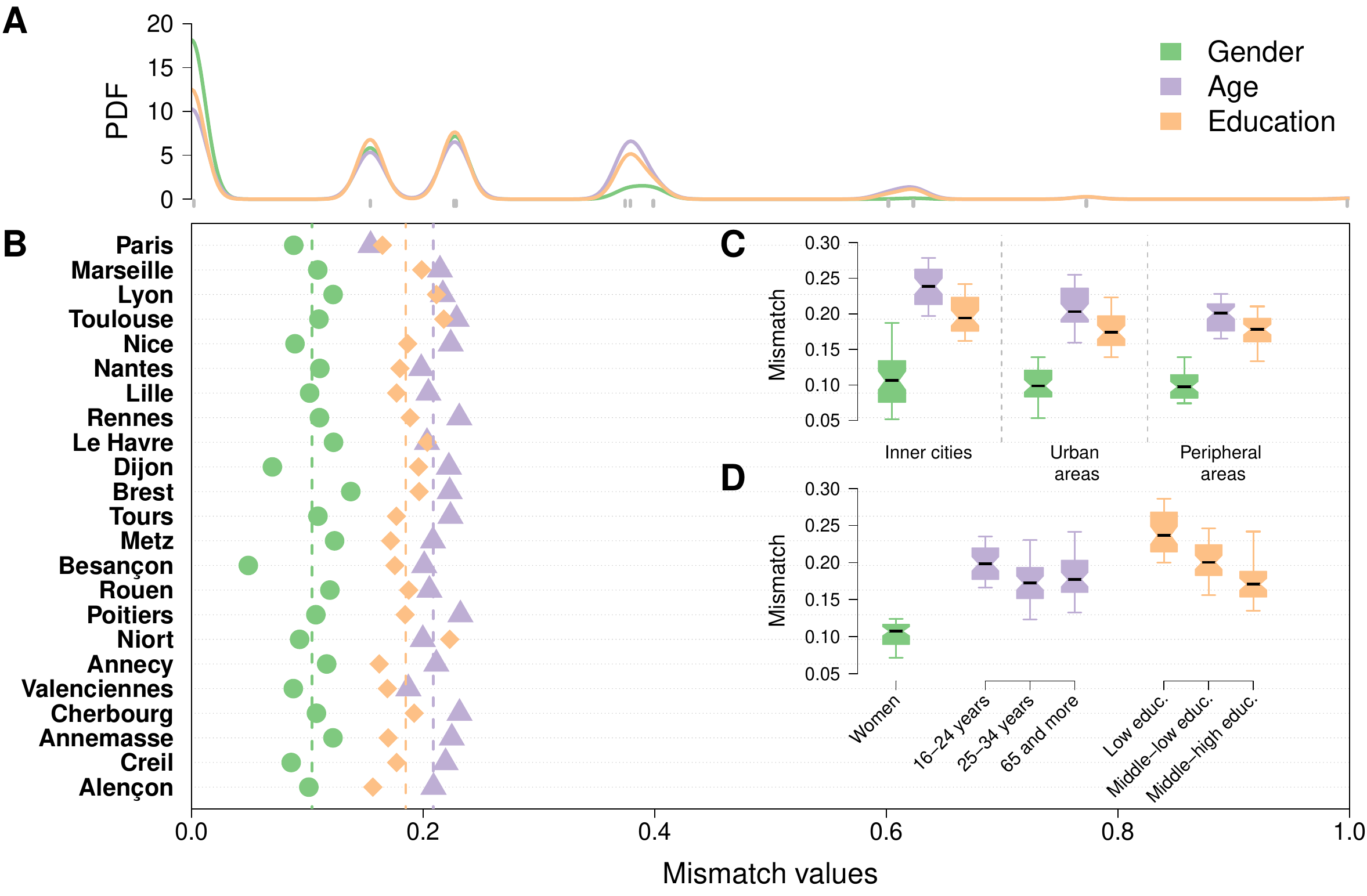}
	\caption{\textbf{Mismatch in district hourly profiles \emph{across} gender, age and education.} (A) Probability Density Function (PDF) of the mismatch according to the sociodemographic variable. A rug plot has been included (grey tick marks at the bottom of the plot) to show the position of the 11 possible mismatch values. (B) Average mismatch for each of the three sociodemographic variables according to the city (only 23 city regions in mainland France are displayed here: the six more populated and also those with the more recent data). The colored dashed lines represent the mismatch values obtained with the 49 city regions taken together. (C) Notched boxplots of the average mismatch per city region for each of the three sociodemographic variables according to the urban gradient. (D) Notched boxplots of the average mismatch per city region for every social group with respect to the reference category (men for the gender, the 35-64 years old category for the age and the high educational level for the education). Each notched boxplot is composed of the first decile, the first quartile, the median, the third quartile and the ninth decile.
		\label{Fig4}}
\end{figure*}

There are no systematic differences in the distribution of district hourly profiles across city regions (Figure \ref{Fig3}C): significant differences are found for 26 city regions among the 49 under consideration (53\%) when compared their distribution to the whole distribution in the remaining 48 cities (refer to Figure S3 in Supplementary Material to know the 26 city regions with significant differences). At a more local scale, we observe large and significant differences in district hourly profiles according to urban gradient as shown in Figure \ref{Fig3}D (see also Figure S3 in Supplementary Material for percentages of sociodistricts belonging to each profile according to the urban gradient for the 49 city regions). As expected, the two attractive daytime profiles (profiles 1 and 2) concern more frequently areas close to core center and the two attractive nighttime profiles (profiles 4 and 5) concern conversely more frequently areas far to core center.

\subsection* {Mismatch in district hourly profiles : variations across gender, age and education}

This first exploration of the space-time rhythms highlights large variations in district hourly profiles within each of the three sociodemographic variables (cf. Figure \ref{Fig3}B). When exploring mismatch values across gender, across age groups and across educational groups, we observe that about 46\% of districts have a gender-based mismatch, 96\% at least one age-based mismatch, 94\% at least one educational-based mismatch and finally 99\% at least one mismatch among the 13 mismatch values under consideration.

Mismatch values are higher for age than for education, and even more than for gender (Figure \ref{Fig4}B). While mismatch values are broadly similar when comparing city regions, they vary greatly according to the urban gradient: mismatch values across age and across education groups were found to be significantly higher in inner cities than in urban or peripheral areas (see notched boxplots in Figure \ref{Fig4}C). 

Finally, with respect to districts' global attractiveness over the 24 hour period (from profiles obtained with the 'All' category) mismatch values both for age and education are widened in districts with large increase of population during the day (profile 1), and to a lesser extent in districts with large decrease of population during the day (profile 5). Similar pattern is also apparent for gender-based mismatch, but in a less marked way (Figure S4 in Supplementary Material).

When we compare gender-based mismatch, age-based mismatch and education-based mismatch using men, middle-age and high educated people respectively as a reference (Figure \ref{Fig4}D), we observe that the largest mismatch value concerns the low educated group (vs. high educated) while the smallest value is observed for women (vs. men). About age-based mismatch, the largest mismatch value concerns the youngest group (vs. 35-64 yrs.) A progressive gradient appears between educational subgroups since mismatch' median values are found to significantly decrease as educational level increases.

\begin{figure*}[!ht]
	\centering 
	\includegraphics[width=\linewidth]{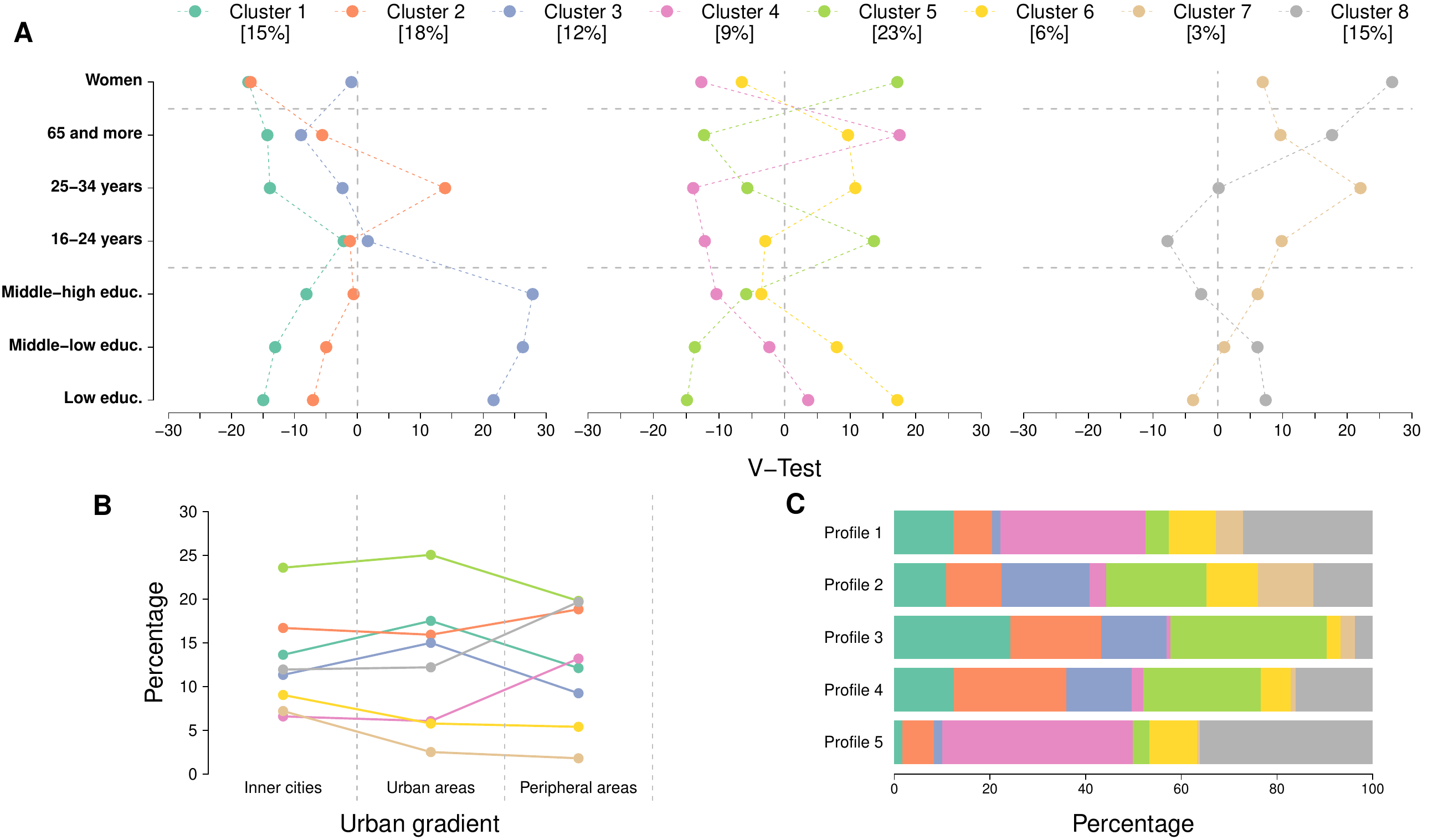}
	\caption{\textbf{Cluster of districts based on mismatch values.} (A) Comparison of the average mismatch values across the three sociodemographic variables per cluster with respect to the average over all districts. (B) Percentage of districts per cluster according to the urban gradient. (C) Percentage of districts per cluster according to the profile obtained with the 'All' category (total population).
		\label{Fig5}}
\end{figure*}

\subsection*{Cluster of districts based on mismatch in their hourly profiles}

To combine all together gender-based mismatch, age-based mismatch and education-based mismatch, we clustered districts exhibiting a similar set of mismatch values in their hourly profiles (see Figure S5 and Table S2 in Supplementary Material for more information about clustering outputs). Eight clusters of districts emerged (Figure \ref{Fig5}A). Cluster 1 is the only one cluster with districts combining all together low mismatch in their hourly profiles for the seven mismatch values by comparison with the average mismatch in the whole sample. Clusters 2, 3 and 4 also combine districts with low mismatch in their hourly profiles but only for two among the three sociodemographic variables: large mismatch values are found for the middle-age group (25-34 yrs.) in cluster 2, for the three educational groups in cluster 3, and for elderly group (65 yrs. and more) in cluster 4. Clusters 5 and 6 gather districts combining large mismatch values for two among the three sociodemographic variables: there are large mismatches in district hourly profiles for gender and age (16-24 yrs.) in cluster 5, and for age and education in cluster 6. Finally, clusters 7 and 8 gather districts combining large mismatch values all together for the three sociodemographic variables. While districts in cluster 7 exhibit large mismatch values for gender, the three age groups and the middle-high educational group, districts in cluster 8 exhibit more specifically large mismatch values for women, elderly people and middle-low and low educated people (i.e. the groups we have defined as belonging to non-dominant groups).

According to the 49 city regions (Figure S6 in Supplementary Material), percentages of districts belonging to cluster 1 vary from 0\% (Angers Region) to 36\% (Longwy Region) and percentages of districts belonging to cluster 8 vary from 0\% (Fort-de-France, Martinique) to 40\% (Angoulême Region).

Cluster distributions highlight large and significant differences according to urban gradient and the five profiles. Districts combining all together low mismatch in their hourly profiles for the seven mismatch values (cluster 1) are more frequent in the urban areas than in inner cities or in peripheral areas (Figure \ref{Fig5}B) and are widely over-represented in areas without daily variation (see profile 3 in Figure \ref{Fig5}C). We also note that districts with large mismatch in their hourly profiles for the youngest group are under-represented in peripheral areas, compared to inner cities and urban areas (see clusters 5 and 7 in Figure \ref{Fig5}B): they are almost absent from very attractive 'nighttime' areas (see profile 5 in Figure \ref{Fig5}C). At the other end of the spectrum, districts exhibiting all together large mismatch in hourly profiles for women, elderly people and low educated people (cluster 8) are similarly present in inner cities and in urban areas but are markedly more frequent in peripheral areas (Figure \ref{Fig5}B). Actually they are widely over-represented in very attractive 'daytime' and 'nighttime' areas (profiles 1 and 5) compared to stable profile (profile 3). This result deserves to be linked with a previous finding showing that gender-based mismatch, age-based mismatch and education-based mismatch considered separately were larger in inner cities and in very attractive 'daytime' areas. It means that the geography of large differences in district hourly profiles change when mismatch is considered in a combined way or separately for gender, age or education. Adopting an intersectional approach crossing gender, age and education makes it possible to emphasize that mismatch in district hourly profiles across social groups is not exclusive to attractive central cities but also concerns peripheral areas with a large drop in their daytime population.

\begin{table*}
	\begin{center}
		\caption {Average mismatch (and standard deviation) between district hourly profiles of men, middle-age and high educated people ('Dominant groups') and of women, elderly and low educated people ('Non-dominant groups')}
		\label{Tab}
		\begin{adjustbox}{width=11 cm}
			\begin{tabular}{llll}
				\\
				\hline
				& \textbf{\#Districts} & \textbf{'Dominant groups'} & \textbf{'Non-dominant groups'} \\ \hline
				
				All & \textit{2561} & 0.127 (0.116) & 0.131 (0.109)\\ \hline
				
				\textit{Per urban gradient} & \\
				Inner cities & \textit{653} & 0.157 (0.127) & 0.145 (0.121)\\
				Urban areas & \textit{1074} & 0.123 (0.111) & 0.121 (0.107)\\
				Peripheral areas & \textit{834} & 0.107 (0.108) & 0.132 (0.1)\\
				\hline
				
				\textit{Per profiles} & \\ 
				Profile 1 & \textit{122} & 0.104 (0.146) & 0.257 (0.133)\\
				Profile 2 & \textit{397} & 0.178 (0.131) & 0.152 (0.105)\\
				Profile 3 & \textit{878} & 0.13 (0.113) & 0.076 (0.091)\\
				Profile 4 & \textit{813} & 0.13 (0.098) & 0.123 (0.087)\\
				Profile 5 & \textit{351} & 0.059 (0.094) & 0.221 (0.086)\\
				
				\hline
				
			\end{tabular}
		\end{adjustbox}
	\end{center}
\end{table*}

\subsection* {Mismatch in district hourly profiles between dominant groups: similar than between non-dominant groups?}

To complete the analysis of space-time profiles with an intersectional lens, we finally computed mismatch in district hourly profiles \emph{between} the three dominant groups: men, middle-age and high educated people. For comparison purposes, we also computed the mismatch \textit{between} three non-dominant groups: women, elderly and low educated people. Mismatch in hourly profiles appear for about 68\% of districts when comparing profiles of the three dominant groups and for about 75\% of districts when comparing profiles of the three non-dominant groups. For the whole districts (Table \ref{Tab}), average mismatch values are almost equivalent between dominant groups (0.127) and between non-dominant groups (0.131).

According to the urban gradient (Table \ref{Tab}), the largest mismatch values are both found in inner cities in comparison with urban areas and peripheral areas. Dominants' mismatch and non-dominants' mismatch are found to have equivalent values in inner cities and in urban areas. But, dominants' mismatch is widely smaller than non-dominants' mismatch in peripheral areas (Table \ref{Tab}). We also observe an interesting pattern when comparing dominants' mismatch and non-dominants' mismatch according to districts global attractiveness over the 24-hour period (from profiles obtained with the 'All' category) (Table \ref{Tab}). While dominants' mismatch and non-dominants' mismatch are found to be roughly equivalent in districts belonging to profile 2 (small increase of total population during the day) and to profile 4 (small decrease of total population during the day), dominants' mismatch is widely smaller than non-dominants' mismatch in districts with large increase (profile 1) or large decreases (profile 5) of total population during the day and conversely dominants' mismatch is much larger than non-dominants' mismatch in districts with stable temporal profile (profile 3).

\subsection* {Methodological discussion}

We based the present analysis on the intersection of three structural dimensions (gender, age and education level) assuming in accordance with Bilge \citep{Bilge_2010} that it is appropriate to treat intersectionality as a meta-principle requiring to be adjusted and rounded out in respect of the particular fields of study and research aims, and to accept the multiplicity of its empirical usages and of its theoretical influences. We did not consider ethnicity or race because origin-destination surveys have not collected this information. This may constitute an unfortunate limitation given the racial segregation and the anti-foreigners discrimination that could be observed in France during both day and night. Ethnic home-based segregation has indeed been enhanced in the Paris region \citep{Preteceille_2011} and the whole French territory \citep{PanKeShon_2010} from census data on nationality and country of birth. A more qualitative study also described anti-foreigners discrimination occurring in French privileged and inner cities areas where foreigners (or considered as such) are present during the day \citep{Najib_2021}. This study echoes our research by pointing out that the perpetrators of racial or religious discrimination feel comfortable with taking action in these areas precisely because 'they feel themselves to be in the majority or dominant' \citep{Najib_2021}. We have also excluded deliberately some other available factors such as occupational status or access to car. These two factors are well-known in time-geography literature to impact space-time constraints and travel patterns and to be unequally distributed across gender, age and education subgroups \citep{Kwan_1999}. From our sample, we observed for example that among middle-age population (35-64 years), fraction of employed people is largely smaller for women than for men (68\% vs. 77\%) and progressively increase with educational level (from 41\% to 83\%). Although factors such as occupational status or access to transportation matter to understand why districts exhibit large dissimilarities in their hourly profiles according to gender, age and educational groups, these factors are more an expression of the power relations in contemporary society and then do not deserve to be included as primary structural factors. It would be interesting for further studies to measure the relative weight of these explanatory factors on the intersectional patterns of everyday geographies, but this is beyond the scope of this paper.

In the paper we used aggregated data issued from 49 French origin-destination surveys. These surveys covered geographical areas of varying size various, sometimes up to the department or even a region in the case of the Paris region ('Île-de-France') and were carried out between 2009 and 2019 inducing a time lag of up to 10 years between surveys. In spite of the variability in geographical coverage and the time lag in data collection, we underline a relative concordance between city regions when exploring their district hourly profiles and extent of their mismatch: it is rather the internal differences within city regions according to urban gradient or population variations across the 24-hour period that stand out. In addition to the year itself, the duration and the year period of the data collection could also impact comparability between city regions. Actually, the 49 surveys were carried out over a maximum period of 20 weeks between October and April (excluding school holidays) and over a minimum period of 8 weeks to smooth exceptional (weather) events \citep{Cerema}. Everyday geographies which are studied here only concern weekdays (Monday-Friday) and districts hourly profiles would probably largely differ if they were derived from daily trips on Saturdays or Sundays. Finally, as all the daily mobility data were collected before the Covid-19 health pandemic, it would be appropriate to renew the analysis in order to consider the spatial and time restrictions on daily trips from March 2020 and the associated changes in gender-based, age-based and educational-based hourly locations within cities.

As for any spatial analysis, our findings are dependant both of the scale and of areal partitioning of our primary spatial units, i.e. the districts. From data collected in the origin-destination surveys, districts are used to aggregate for every hour the (weighted) number of ambient respondents. The 2,572 studied districts are quite heterogeneous in terms of size and residential population (see Table S1 in Supplementary Material for summary statistics). This heterogeneity is a direct reflection of variations in population densities within city regions. There is no choice but to use heterogeneous spatial units when one wants to ensure sufficient sample size for statistical analysis and to protect confidentiality of personal data for the provision of open data. Moreover, this spatial heterogeneity is not specific to our survey data: analysis from mobile phone data are also very dependant of heterogeneous densities of mobile phone towers in all countries.

Rural areas are under-represented in our district sample (see Figure \ref{Fig1}) because of the geographical coverage of origin-destination surveys \citep{Cerema}. French areas belonging to 'diagonal of low densities'  -  historically more aged and more socially deprived that other parts of the French territory \citep{Pistre_2010} - are then excluded from the present analysis. It could be interesting to find empirical data allowing to extend our analysis to remote French rural areas but also to other countries to see whether (and where) district hourly profiles of dominant groups are also found to diverge from those of the others population groups.

Finally we have developed a two-steps methodological process to group together districts exhibiting similar hourly signals and to measure afterwards mismatch in district hourly profiles according to social groups. It is important to note that the main purpose of the first step is to reduce the noise in order to obtain characteristic profiles. It is also worth emphasizing that the proposed method could be used to compare various kind of space-time profiles, and not exclusively district profiles over the 24 hour period. It could for example be used to compare mismatch in district \textit{yearly} profiles according to social groups and to better grasp the intersectional forms of population concentration per district over the years resulting from diverse residential mobility patterns.

\section*{Conclusion}

When exploring hourly rhythms of ambient populations within cities, five district hourly profiles (two 'daytime attractive', two 'nighttime attractive' and one more 'stable') emerged: they are broadly distributed in a similar way across the 49 French city regions but unequally distributed according to social groups and to urban gradient.

Districts exhibit large dissimilarities in their hourly profiles according to educational and age groups, and to a lesser extent according to gender. Three numbers can sum up the situation: about 46\% of districts are found to have a gender-based mismatch in their hourly profiles, 96\% at least one mismatch across the four age groups and 94\% at least one mismatch across the four educational groups. By highlighting that district hourly profiles of different social groups rarely overlap (i.e. exhibit a mismatch), this paper provides an argument for the importance of disaggregating ambient populations according to their demographic or social profiles. By comparing the geographical locations of social groups over the 24-hour period, the present paper gives more broadly insight towards the uneven geography of daily destinations (not only jobs, but also shops, leisure places, etc.) and their unequal use according to gender, age and social groups. 

Moreover, some districts have the particular feature of combining all together large mismatch in hourly profiles across gender (women vs. men), age (elderly vs. middle-age people) and education (low educated vs. high educated people). These districts are spread across the French city regions but are over-represented both in the areas with large increase or large decrease of population during the day. While recent quantitative research exploring daytime population variations mainly focus on inner cities and on single social dimension, the present paper highlights that the space-time dissimilarities extend outside inner cities areas and concern mutually constitutive forms of exclusion. It thus provides an awareness of the intersectional dimension that spatial segregation can take over a 24-hour period.

Finally, synchronization of hourly rhythms is found to differ according to population segments under consideration but also according to areas and their global attractiveness over the 24 hour period. In areas exhibiting large increase or large decrease of population during the day, we note that hourly rhythms of men, middle-age and high educated people (dominant groups) are widely more synchronous to each other than rhythms of women, elderly and low educated people (non-dominant groups). Conversely, in areas with stable people concentration over the 24 hour period, divergence in hourly profiles synchronization between these population segments is smaller (and ordered in a reverse way) than in very attractive 'daytime' and 'nighttime' areas: rhythms between non-dominant groups are found to be slightly more synchronized than rhythms between dominant groups. Here again, the present paper underlines that area hourly rhythms, and especially their synchronization, gain to be explored over the whole territory and across different sociodemographic attributes.

Unlike the rare studies about everyday segregation mainly limited to one city region, the present paper concerns a large sample of city regions allowing to observe a relative concordance between city regions when exploring their district hourly profiles and their mismatch values across social groups and to emphasize conversely large internal differences within city regions according to districts' global attractiveness over the 24 hour period rather than according to the sole urban gradient.

Through this original analysis of the intersectional forms of everyday geographies, we extend the scope of segregation literature mainly centered on home places from a single social dimension. Literature has underlined for decades that spatial aggregation in residential areas both reflects and structures power relations and inequalities: residential segregation prompts upper social classes to accumulate resources and capital and to perpetuate their differential access to privileges \citep{Pincon_2018} and at the other end of the social hierarchy, makes poor people much more vulnerable and more ’disadvantaged’ \citep{Wilson_1987}. The present research is a first step to broaden the scope of these residential-based mechanisms and to enhance knowledge of space-time (de)synchronization across gender, age and educational groups: it sheds new light on areas where peers are synchronously located over the 24-hour period and thus potentially in better position to interact and to defend their common interests. It may feed future intersectional studies dedicated to everyday geography and to its effect on the dynamics of power relations and inequality.

\vspace*{0.5cm}
\section*{Acknowledgments}

We thank the Mobiliscope team (in particular Aurélie Douet and Guillaume Le Roux), its partners and funders (CNRS, ANCT, Ined and Labex DynamiTe) as well as the French data producers (Cerema and DRIEA) and their main distributor (Progedo-Adisp). We thank also members of the collective Eighties for the stimulating exchanges.

\bibliographystyle{myapalike}
\small
\bibliography{Intersectionality}

\onecolumngrid

\makeatletter
\renewcommand{\fnum@figure}{\sf\textbf{\figurename~\textbf{S}\textbf{\thefigure}}}
\renewcommand{\fnum@table}{\sf\textbf{\tablename~\textbf{S}\textbf{\thetable}}}
\makeatother

\setcounter{figure}{0}
\setcounter{table}{0}
\setcounter{equation}{0}

\newpage
\clearpage
\newpage
\section*{Appendix}

\section*{Supplementary figures}

\begin{figure}[!ht]
	\centering 
	\includegraphics[width=12cm]{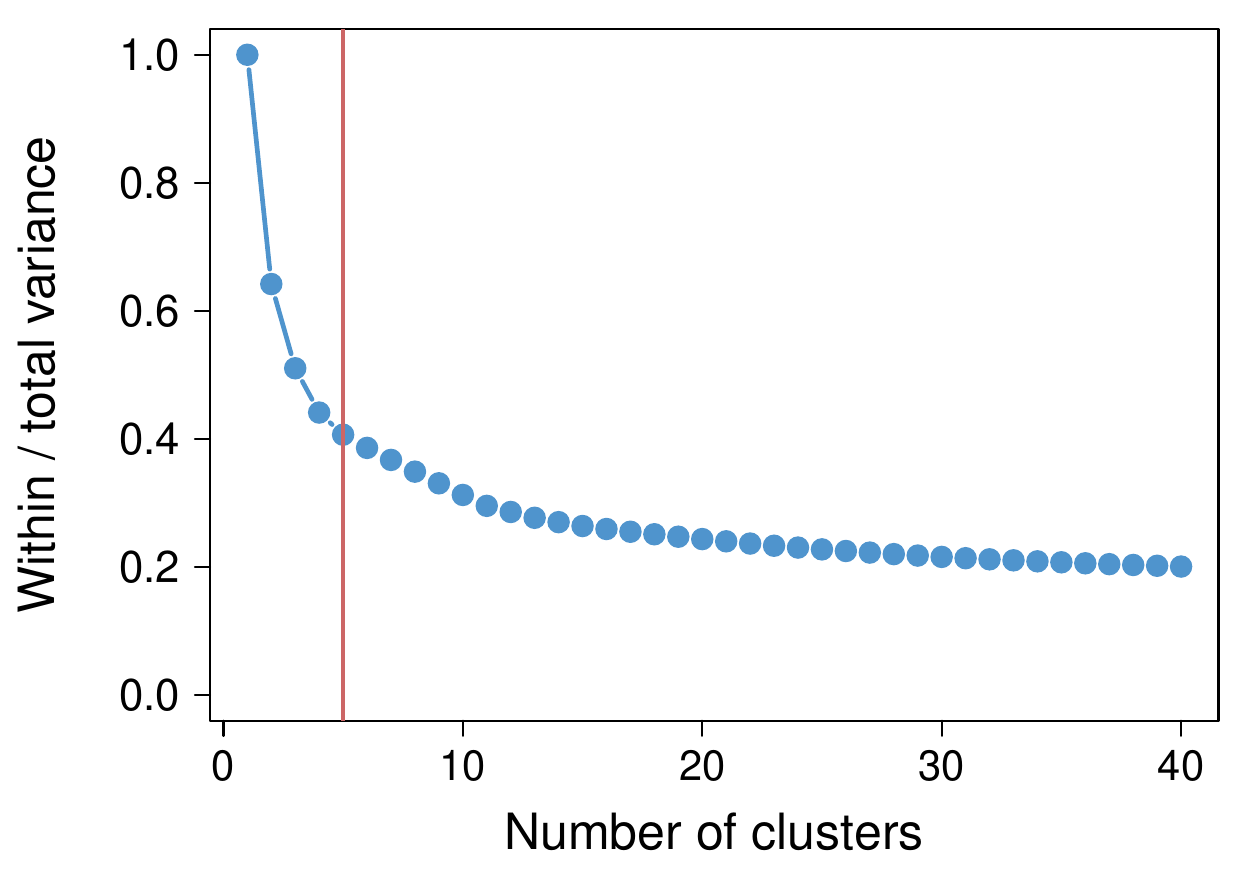}
	\caption{\textbf{Ratio between the within-group variance and the total variance as a function of the number of clusters.} We performed an ascending hierarchical clustering using Ward's metric and Euclidean distances as agglomeration method and dissimilarity metric to cluster the 28,281 sociodistricts according to their hourly signals. We identified five main hourly profiles. 
		\label{FigS1}}
\end{figure}

\begin{figure}[!ht]
	\centering 
	\includegraphics[width=\linewidth]{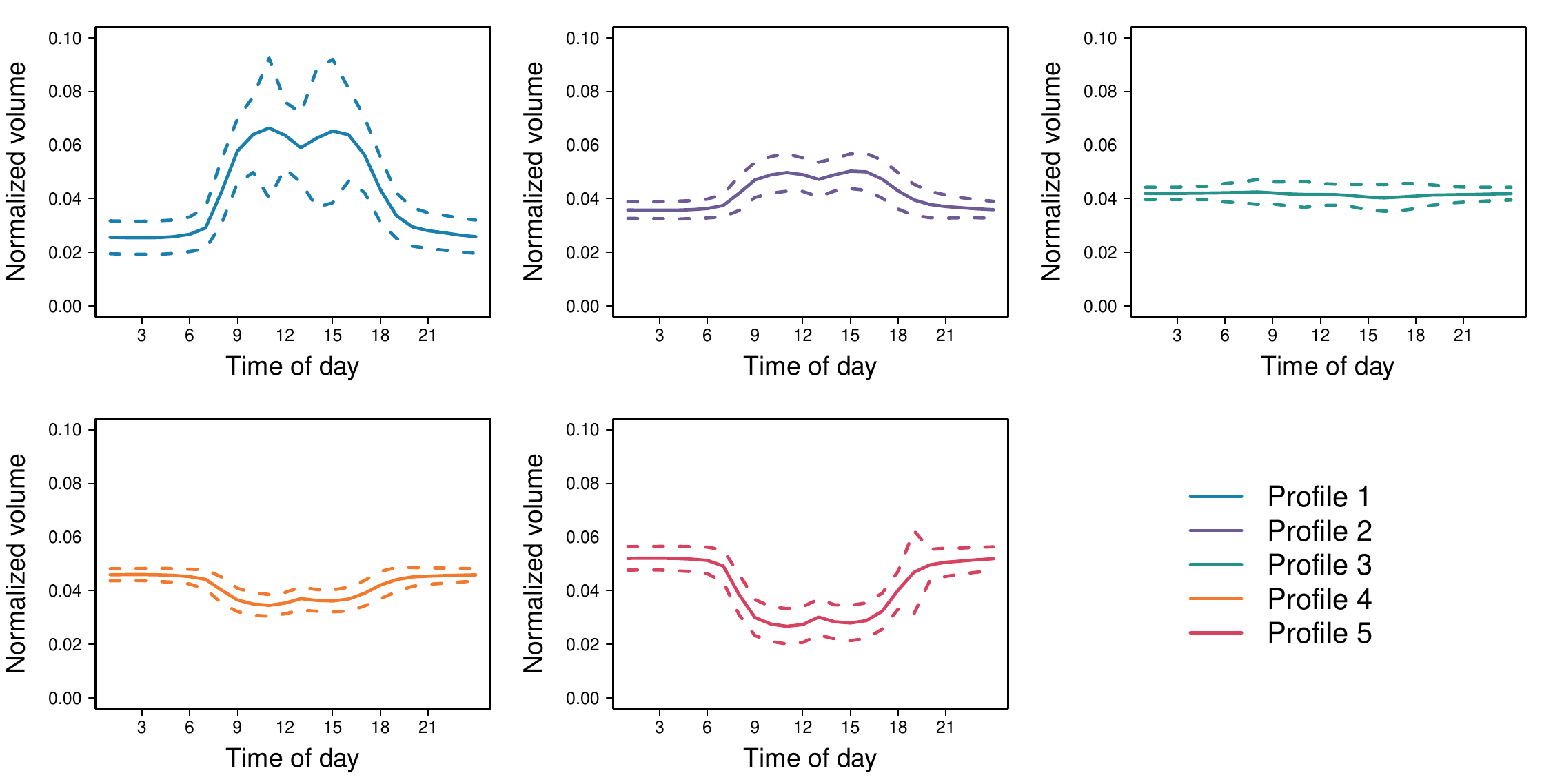}
	\caption{\textbf{Hourly profiles.} Average hourly signals representing the five profiles. The solid lines represent the average normalized volume, while the dashed lines represent one standard deviation. 
		\label{FigS2}}
\end{figure}

\newpage
\begin{figure}[!ht]
	\centering 
	\includegraphics[width=\linewidth]{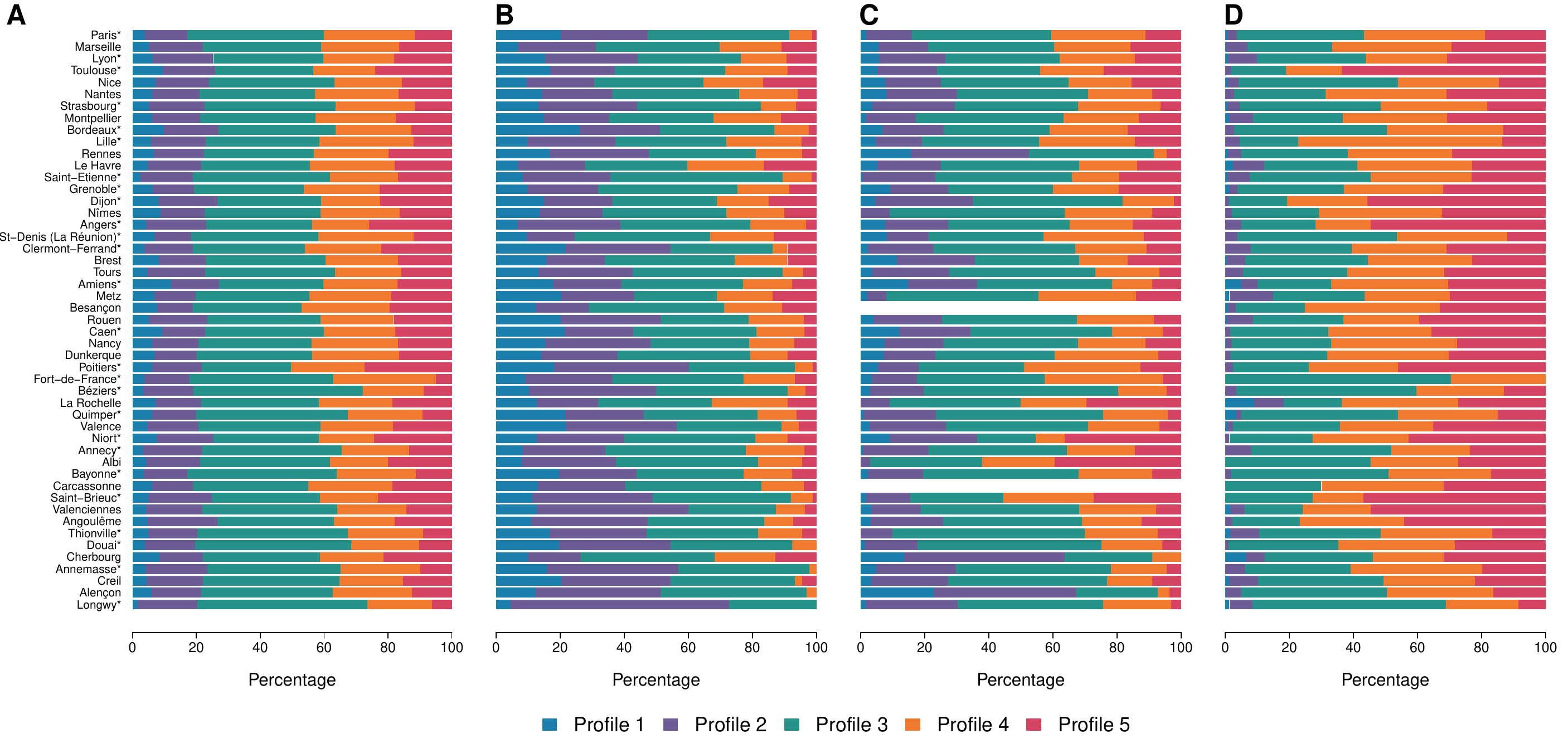}
	\caption{\textbf{Percentage of sociodistricts belonging to each profile according to the city region and the urban gradient.} (A) All districts. (B) Inner cities. (C) Urban areas. (D) Peripheral areas.\\
		\textit{Note: * indicates city regions that exhibit a significant difference in the distribution of district hourly profiles when compared to the whole distribution of the remaining 48 cities (from chi-square test; $p<0.01$).}
		\label{FigS3}}
\end{figure}

\begin{figure}[!ht]
	\centering 
	\includegraphics[width=\linewidth]{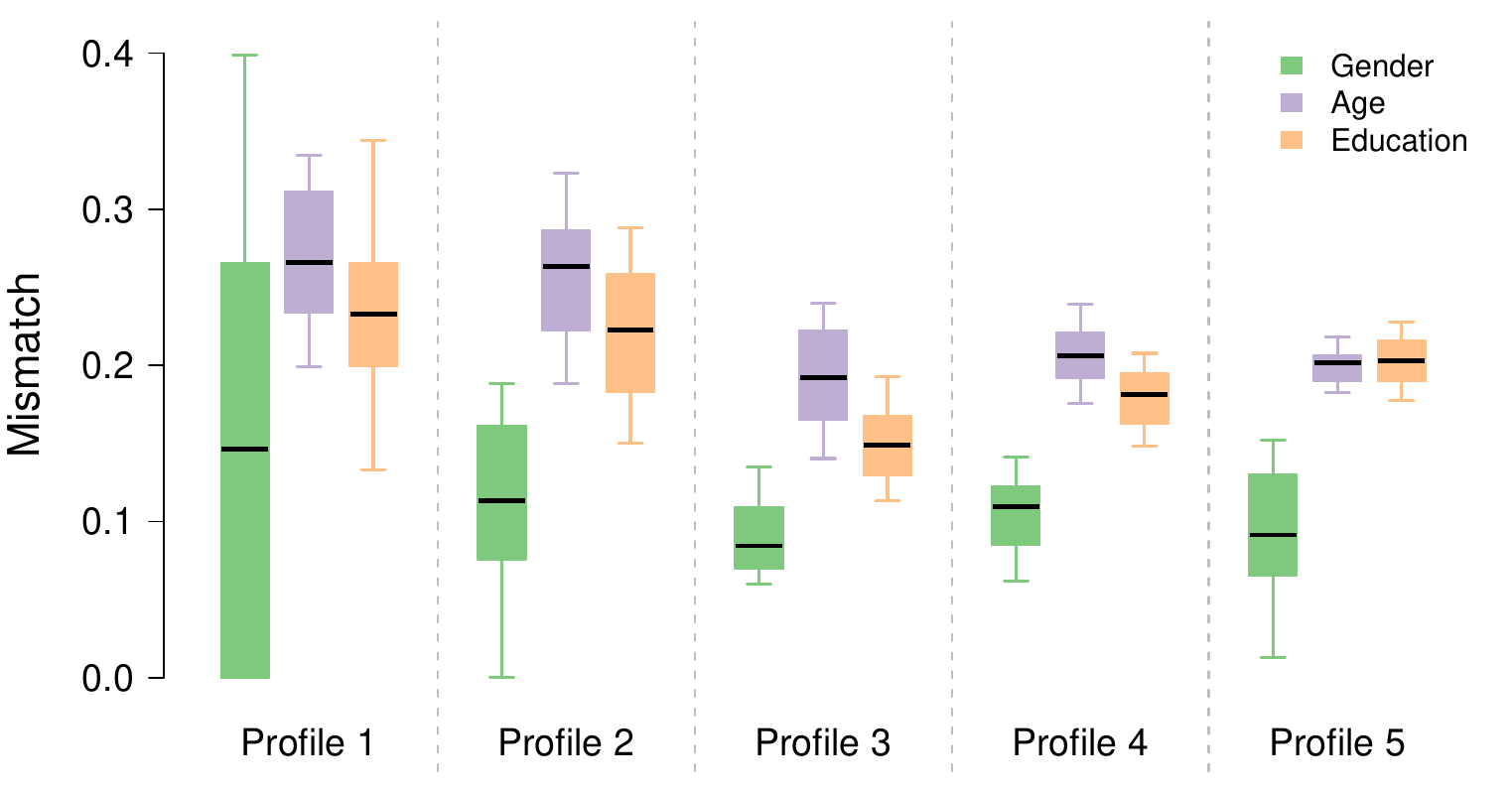}
	\caption{\textbf{Mismatch within each of the three sociodemographic variables according to the profile.} Boxplots of the average mismatch per city according to the sociodemographic variable and the profile obtained with the 'All' category (total population). Each boxplot is composed of the first decile, the first quartile, the median, the third quartile and the ninth decile. 
		\label{FigS4}}
\end{figure}

\newpage
\begin{figure}[!ht]
	\centering 
	\includegraphics[width=12cm]{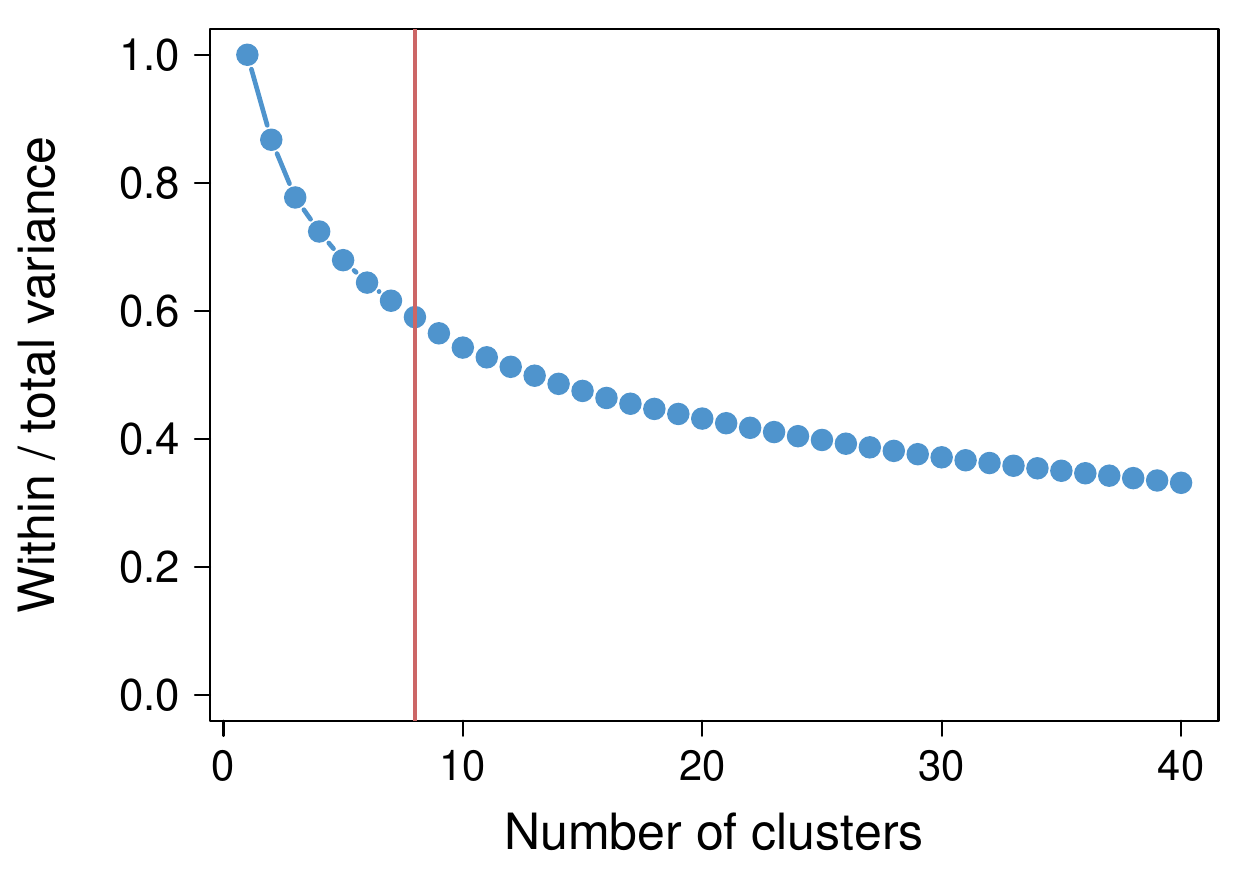}
	\caption{\textbf{Ratio between the within-group variance and the total variance as a function of the number of clusters.} We performed an ascending hierarchical clustering using Ward's metric and Euclidean distances as agglomeration method and dissimilarity metric to cluster the 2,561 districts (after removing the 11 districts lacking of information, i.e. sociodistricts with no activity) based on the 7 within-mismatch values. We identified eight clusters. 
		\label{FigS5}}
\end{figure}

\newpage
\begin{figure}[!ht]
	\centering 
	\includegraphics[width=12cm]{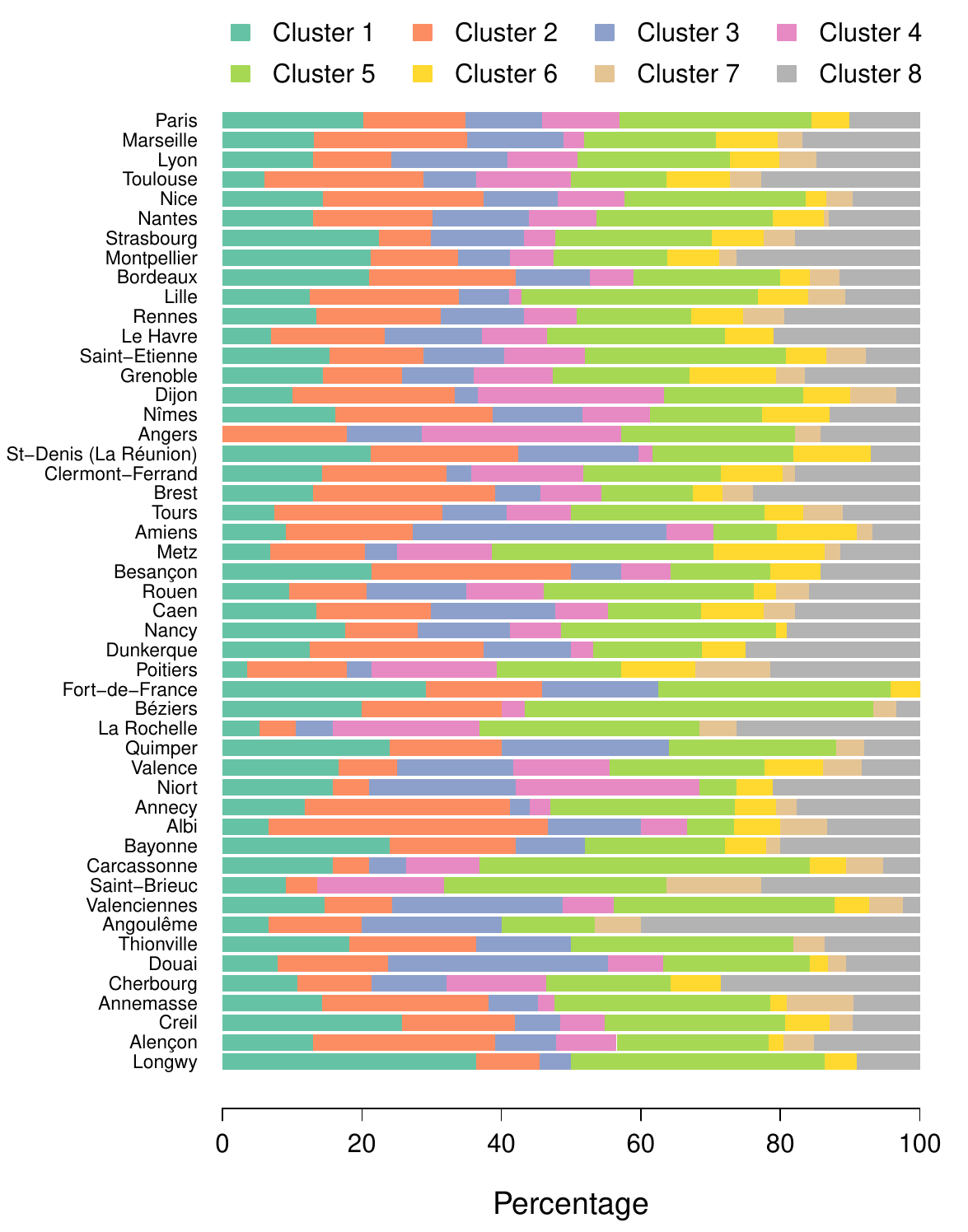}
	\caption{\textbf{Percentage of districts per cluster according to the city region.}
		\label{FigS6}}
\end{figure}

\newpage
\section*{Supplementary tables}

\begin{table}[!ht]
	\caption{\textbf{Summary information about the 49 French origin-destination surveys.}
		\label{TabS1}}
	
	\begin{center}
		\begin{adjustbox}{max width=\textwidth}
			\begin{tabular}{llllllll}
				\hline
				City region & Year & \begin{tabular}[c]{@{}l@{}}Number of\\respondents\\(16 yrs.and +)\end{tabular} & \begin{tabular}[c]{@{}l@{}}Number of\\districts\end{tabular} & \begin{tabular}[c]{@{}l@{}}District area:\\median (min-max)\\in km²\end{tabular} & \begin{tabular}[c]{@{}l@{}}Number of\\respondents per\\district of residence:\\median (min-max)\end{tabular} & \begin{tabular}[c]{@{}l@{}}Residential\\population\\(2013 census) \end{tabular} & \begin{tabular}[c]{@{}l@{}}Residential\\population in\\inner cities\\(2013 census)\end{tabular} \\ \hline
				Paris & 2010 & 26,312 & 109 & 26 (3-1324) & 234 (128-419) & 11,959,800 & 2,229,600 \\
				Marseille & 2009 & 19,380 & 137 & 18 (0.3-342) & 141 (116-161) & 2,006,000 & 855,400 \\
				Lyon & 2015 & 24,072 & 169 & 10 (0.4-346) & 140 (124-254) & 2,388,400 & 500,700 \\
				Toulouse & 2013 & 11,141 & 66 & 11 (0.9-237) & 156 (130-326) & 1,087,200 & 458,300 \\
				Nice & 2009 & 14,989 & 104 & 5 (0.4-505) & 143 (127-163) & 1,104,300 & 342,300 \\
				Nantes & 2015 & 17,358 & 123 & 21 (0.6-331) & 139 (113-195) & 1,340,600 & 292,700 \\
				Strasbourg & 2009 & 10,052 & 67 & 35 (0.6-581) & 144 (128-228) & 1,109,500 & 275,700 \\
				Montpellier & 2014 & 11,433 & 80 & 10 (0.3-547) & 140 (127-240) & 791,000 & 272,100 \\
				Bordeaux & 2009 & 13,793 & 95 & 10 (0.4-1206) & 143 (125-218) & 1,505,500 & 243,600 \\
				Lille & 2016 & 7,950 & 57 & 5 (1-56) & 138 (126-157) & 1,129,100 & 231,500 \\
				Rennes & 2018 & 9,317 & 68 & 55 (0.8-466) & 137 (118-155) & 1,094,300 & 211,400 \\
				Le Havre & 2018 & 6,540 & 43 & 25 (0.7-256) & 143 (129-263) & 502,800 & 172,100 \\
				Saint-Etienne & 2010 & 8,525 & 52 & 30 (1.2-345) & 156 (139-224) & 598,800 & 172,000 \\
				Grenoble & 2010 & 13,834 & 97 & 14 (0.2-834) & 140 (124-278) & 816,000 & 160,200 \\
				Dijon & 2016 & 4,372 & 30 & 6 (0.7-241) & 144 (128-180) & 314,300 & 153,000 \\
				Nîmes & 2015 & 4,519 & 31 & 8 (0.3-152) & 144 (134-166) & 283,800 & 150,600 \\
				Angers & 2012 & 3,983 & 28 & 17 (2-111) & 140 (126-161) & 317,800 & 150,100 \\
				St-Denis (La Réunion)  & 2016 & 13,801 & 99 & 11 (0.6-275) & 139 (122-161) & 835,100 & 142,400 \\
				Clermont-Ferrand & 2012 & 8,052 & 56 & 39 (0.7-471) & 143 (132-159) & 667,900 & 141,500 \\
				Brest & 2018 & 6,519 & 46 & 13 (0.9-236) & 141 (125-162) & 415,000 & 139,400 \\
				Tours & 2019 & 7,624 & 54 & 43 (0.5-637) & 141 (123-154) & 600,300 & 134,800 \\
				Amiens & 2010 & 7,097 & 45 & 13 (0.8-410) & 147 (129-238) & 338,800 & 132,700 \\
				Metz & 2017 & 6,492 & 44 & 13 (1.1-176) & 142 (126-246) & 385,800 & 118,600 \\
				Besançon & 2018 & 3,980 & 28 & 5 (0.5-191) & 142 (129-156) & 205,800 & 117,000 \\
				Rouen & 2017 & 8,905 & 63 & 15 (1.1-231) & 142 (121-156) & 726,700 & 110,800 \\
				Caen & 2011 & 10,000 & 67 & 15 (0.5-647) & 151 (127-167) & 689,900 & 107,200 \\
				Nancy & 2013 & 9,657 & 68 & 17 (0.6-407) & 142 (126-159) & 570,700 & 104,100 \\
				Dunkerque & 2015 & 4,537 & 32 & 9 (1.1-117) & 143 (129-153) & 266,100 & 89,900 \\
				Poitiers & 2018 & 4,106 & 28 & 51 (1.1-228) & 140 (129-255) & 238,900 & 87,400 \\
				Fort-de-France & 2014 & 4,179 & 24 & 36 (4.7-129) & 156 (129-283) & 385,600 & 84,200 \\
				Béziers & 2014 & 4,212 & 30 & 52 (1.2-822) & 139 (130-153) & 301,300 & 74,800 \\
				La Rochelle & 2011 & 2,902 & 19 & 8 (1.3-40) & 153 (140-165) & 148,500 & 74,300 \\
				Quimper & 2013 & 4,696 & 30 & 57 (2.5-260) & 152 (144-196) & 336,300 & 63,500 \\
				Valence & 2014 & 5,143 & 36 & 41 (1.3-477) & 142 (126-156) & 358,000 & 61,800 \\
				Niort & 2016 & 2,869 & 19 & 15 (1.1-156) & 151 (145-159) & 118,300 & 57,400 \\
				Annecy & 2017 & 4,953 & 34 & 27 (0.9-486) & 144 (132-172) & 347,600 & 52,000 \\
				Albi & 2011 & 2,339 & 15 & 7 (1.3-54) & 157 (147-160) & 81,400 & 49,300 \\
				Bayonne & 2010 & 7,387 & 50 & 20 (0.8-642) & 148 (132-171) & 385,400 & 47,500 \\
				Carcassonne & 2015 & 2,891 & 19 & 36 (0.8-246) & 153 (144-160) & 112,600 & 46,700 \\
				Saint-Brieuc & 2012 & 3,570 & 22 & 8 (1.1-35) & 154 (149-248) & 115,900 & 45,300 \\
				Valenciennes & 2019 & 5,647 & 41 & 12 (0.9-51) & 137 (127-151) & 348,600 & 42,900 \\
				Angoulême & 2012 & 2,684 & 18 & 14 (1.5-151) & 149 (137-174) & 140,600 & 42,000 \\
				Thionville & 2012 & 3,609 & 22 & 7 (0.6-62) & 164 (156-173) & 183,600 & 41,600 \\
				Douai & 2012 & 5,344 & 38 & 8 (1-54) & 140 (126-160) & 255,300 & 41,200 \\
				Cherbourg & 2016 & 4,207 & 28 & 42 (1.3-231) & 150 (143-159) & 205,300 & 37,100 \\
				Annemasse & 2016 & 5,905 & 42 & 36 (0.8-232) & 138 (122-213) & 532,000 & 34,600 \\
				Creil & 2017 & 4,581 & 31 & 16 (1-144) & 147 (142-154) & 248,400 & 34,300 \\
				Alençon & 2018 & 6,520 & 46 & 145 (1.9-435) & 141 (131-150) & 474,800 & 26,400 \\
				Longwy & 2014 & 3,347 & 22 & 32 (2.2-200) & 152 (145-160) & 179,400 & 14,100 \\ \hline
			\end{tabular}
		\end{adjustbox}
	\end{center}
\end{table}

\begin{table}[!ht]
	\caption{\textbf{Average mismatch (and the associated standard deviation) per cluster, taking dominant groups as reference (men for the gender, 35-64 years old for the age and high educational level for the educational level).}}
	\label{TabS2}
	\begin{center}
		\begin{adjustbox}{max width=\textwidth}
			\begin{tabular}{lccccccc}
				\hline
				\centering Cluster & Women & 16-24 years & 25-34 years & 65 and more & Low educ. & Middle-low educ. & Middle-high educ.  \\
				\hline
				
				All & 0.104 (0.126) & 0.197 (0.196) & 0.173 (0.177) & 0.178 (0.167) & 0.248 (0.204) & 0.206 (0.182) & 0.176 (0.186)\\
				Cluster 1 & 0 (0) & 0.177 (0.143) & 0.056 (0.09) & 0.065 (0.094) & 0.104 (0.12) & 0.093 (0.113) & 0.105 (0.117)\\
				Cluster 2 & 0.011 (0.04) & 0.187 (0.156) & 0.281 (0.118) & 0.138 (0.119) & 0.186 (0.135) & 0.167 (0.129) & 0.171 (0.143)\\
				Cluster 3 & 0.098 (0.107) & 0.214 (0.168) & 0.151 (0.171) & 0.099 (0.102) & 0.483 (0.176) & 0.46 (0.181) & 0.451 (0.186)\\
				Cluster 4 & 0 (0) & 0.042 (0.114) & 0.013 (0.053) & 0.368 (0.106) & 0.296 (0.15) & 0.179 (0.151) & 0.051 (0.107)\\
				Cluster 5 & 0.183 (0.09) & 0.293 (0.214) & 0.137 (0.125) & 0.104 (0.112) & 0.138 (0.124) & 0.116 (0.115) & 0.137 (0.131)\\
				Cluster 6 & 0.042 (0.094) & 0.154 (0.204) & 0.317 (0.182) & 0.299 (0.185) & 0.511 (0.208) & 0.315 (0.154) & 0.126 (0.155)\\
				Cluster 7 & 0.195 (0.162) & 0.398 (0.271) & 0.58 (0.175) & 0.347 (0.243) & 0.167 (0.171) & 0.225 (0.217) & 0.295 (0.224)\\
				Cluster 8 & 0.267 (0.088) & 0.124 (0.172) & 0.174 (0.156) & 0.319 (0.137) & 0.321 (0.176) & 0.26 (0.147) & 0.153 (0.171)\\

				\hline
			\end{tabular}
		\end{adjustbox}
	\end{center}
\end{table}

\end{document}